\documentclass[proof]{WileyASNA-v1}
\articletype{The "$6^{th}$ workshop on Compact Steep Spectrum and GHz-Peaked Spectrum Radio sources" proceeding:}
\revised{October 2021}
\accepted{October 2021}
\usepackage{wrapfig}
\usepackage[nameinlink]{cleveref}
\usepackage{graphicx}
\begin{document}
\title{Studies on slow radio transients.}
\author[1]{Aleksandra Wołowska*}
\author[1]{Magdalena Kunert-Bajraszewska}
\author[2,3]{Kunal Mooley}
\author[4]{Aneta Siemiginowska}
\author[5]{Preeti Kharb}
\author[5]{C. H. Ishwara-Chandra}
\author[3]{Gregg Hallinan}
\author[6]{Mariusz Gromadzki}
\author[7]{Dorota Kozie\l-Wierzbowska}
\authormark{Wołowska et al.}
\address[1]{\orgdiv{Institute of Astronomy}, \orgname{Faculty of Physics, Astronomy and Informatics, NCU}, \orgaddress{\state{Grudziądzka 5/7, 87-100, Toruń}, \country{Poland}}}
\address[2]{\orgdiv{National Radio Astronomy Observatory}, \orgname{P.O.\,Box O, Socorro}, \orgaddress{\state{NM 87801}, \country{USA}}}
\address[3]{\orgdiv{Cahill Center for Astronomy}, \orgname{California Institute of Technology}, \orgaddress{\state{Pasadena, CA 91125}, \country{USA}}}
\address[4]{\orgdiv{Center for Astrophysic}, \orgname{Harvard \& Smithsonian}, \orgaddress{\state{Cambridge, MA}, \country{USA}}}
\address[5]{\orgdiv{National Centre for Radio Astrophysics}, \orgname{Tata Institute of Fundamental Research}, \orgaddress{\state{Ganeshkhind, Pune 411007}, \country{India}}}
\address[6]{\orgdiv{Astronomical Observatory}, \orgname{University of Warsaw}, \orgaddress{\state{Al. Ujazdowskie 4, 00-478 Warsaw}, \country{Poland}}}
\address[7]{\orgdiv{Astronomical Observatory}, \orgname{Jagiellonian University}, \orgaddress{\state{ul. Orla 171, 30-244 Krak\'ow}, \country{Poland}}}

\corres{*\email{ola@astro.umk.pl}}

\abstract{
%A recent Caltech-NRAO Stripe 82 Survey (CNSS), among the numerous new discoveries, revealed a group of sources that are absent in the  Faint Images of the Radio Sky at Twenty Centimeters (FIRST) catalog, and therefore represent a new sample of sources with renewed radio activity. 
We present a brief overview of a very extensive studies of the group of active galactic nuclei (AGNs) that transitioned to radio-loud state over the past few decades.
%including the analysis of their morphology, radio and optical spectra, X-ray properties and physical parameters.\\
The sample consists of twelve sources, both quasars and galaxies, showing the characteristics of gigahertz-peaked spectrum (GPS) objects
undergoing relatively rapid changes, due to the evolution of their newly-born radio jets. 
%with the emphasis that the quasar/galaxy division significantly affects the further evolution of the source. 
Discussed objects also show a wide range of physical parameters such as bolometric
luminosity, black hole mass and jet power,
suggesting a great diversity among young active galactic nuclei and their hosts. Furthermore, we introduce a new observational project, the aim of which will be to investigate and gain a more in-depth understanding of the phenomenon of slow radio transients.}

\jnlcitation{\cname{%
\author{Wo\l owska A.}, 
\author{Kunert-Bajraszewska M.},
\author{Mooley K.}} (\cyear{2021}) et al.,
\ctitle{AGNs that transitioned to radio-loud state}, \cjournal{Astron. Nachr. 2021}, \cvol{}.}

\maketitle
\section{Introduction}\label{sec1}
One of the main issues in modern research on active galactic nuclei (AGNs) is their evolutionary process. The proposed model, in which compact radio sources called the gigahertz-peaked spectrum (GPS)
and compact steep spectrum (CSS) objects propagate their jets and enlarge their sizes to then develop 
into a large-scale FRI/FRII type radio galaxies \citep{Fanti, Readhead, Odea97, Snellen2000} appears to be correct in principle. However, statistical studies show that we observe a surplus of compact, not fully developed sources, which in turn suggests that not every young radio 
galaxy will follow the evolutionary path described above. A few possible reasons for this phenomenon have been discussed in the literature. This is the presence of a dense environment suppressing the jets expansion \citep{Mukherjee}, the enhancement of radio emission in young sources due to their jet-environment interaction, resulting in their over-abundance in selected samples \citep{Morganti, Tadhunter}, and a short time scale ($10^4-10^5$ years) episodic activity of radio objects, causing them to alternately vanish and reappear as compact, young looking sources
%where the activity phase lasts for $10^4-10^5$ years due to the radiation pressure instabilities in the accretion disk
\citep{Reynolds, MKB2010, Czerny, me, Silpa}.
The studies of such transient phenomena have been recently performed by us using the high quality Caltech-NRAO Stripe 82 Survey \citep[CNSS;][]{Mooley16, Mooley19}. 
%So far, the number of sky surveys dedicated to studying this phenomenon has been quite limited. This situation changed when the Caltech-NRAO Stripe 82 Survey \citep[CNSS;][]{Mooley16, Mooley19} and  the  Very Large Array Sky Survey \citep[VLASS;][]{Lacy} appeared - the projects dedicated to tracking transient phenomena and introducing a new quality of observations.\\
The CNSS was carried out with the Jansky Very Large Array (VLA) between 2012 December and 2015 May over five epochs at the S band (2-4GHz). The observations covered the Sloan Digital Sky Survey (SDSS) Stripe 82 region of the 270 deg$^2$ and resulted in the discovery, among others, 
%of many variable and transient sources, a large fraction of which were of AGN origin. Further comparison of the results with the Faint Images of the Radio Sky at Twenty Centimeters \citep[FIRST;][]{White} and the VLA survey of the SDSS Southern Equatorial Stripe \citep[VLA-Stripe 82;][]{Hodge}, has revealed 
of transient sources on timescales <20 years, presumably associated with renewed AGN activity. 
%Discovered sources were further screened for criteria such as pointlikeness, signal-to-noise ratio, flux density and a sufficiently large separation from other bright sources.\\
The final selected sample consisted of 12 transient sources (Tab.\ref{basic}) that were studied further by us in order to reveal the origin and nature of this phenomena \citep{MKB2020, me21}.
%with VLBA, VLA, GMRT, XMM-Newton, Chandra X-ray Observatory and SALT with the addition of archival photometric and spectroscopic SDSS data \citep{me21}.
%We would like to present and discuss the results of our comprehensive, further study of these sources, and its impact on currently accepted theories about jet formation and galactic evolution.\\

\begin{center}
\begin{table}[t]
\fontsize{5}{6}
\caption{Basic properties of a selected sample.}
\centering
\begin{tabular}{ c c c c c c }
\hline
Name & ID & z & $3\sigma_{FIRST} $ & $\rm S_{1.4}$& $\rm logL_{1.4}$\\
     &    &   & [mJy]             & [mJy]& [$\rm W~Hz^{-1}$]\\
(1) & (2) & (3) & (4) & (5) & (6) \\
\hline
221650$+$00& G &0.55&0.39&0.52&23.1\\
221812$-$01&$-$&$-$&0.51&1.60&$-$\\
223041$-$00&G &0.84&0.37&0.98&23.8\\
233001$-$00& Q&1.65&0.32&3.45&24.6\\
010733$+$01&G&0.12&0.47&3.03&22.9\\
013815$+$00&Q&0.94&0.32&1.48&24.5\\
015411$-$01&G&0.05&0.45&3.79&22.3\\
020827$-$00&Q&1.34&0.49&1.90&24.2\\
030533$+$00&G&0.42&0.41&0.88&23.4\\
030925$+$01&G&0.04&0.35&3.69&22.1\\
031833$+$00&G&0.40&0.34&0.79&23.2\\
034526$+$00&G&0.45&0.40&1.22&23.5\\
\hline
\hline
\label{basic}
\end{tabular}
\begin{tablenotes}%%[341pt]
Note: (4) the 3$\sigma$ noise level at the location of the source measured in the 1.4 GHz FIRST images in mJy; (5) the latest value of 1.4 GHz flux density measured based on our VLA observations in mJy. Taken from \citet{me21}.
\end{tablenotes}
\end{table}
\end{center}

\section{Transition events from CNSS}\label{sec2}
\begin{center}
\begin{figure*}[ht!]
    \centering
    \includegraphics[scale=0.4]{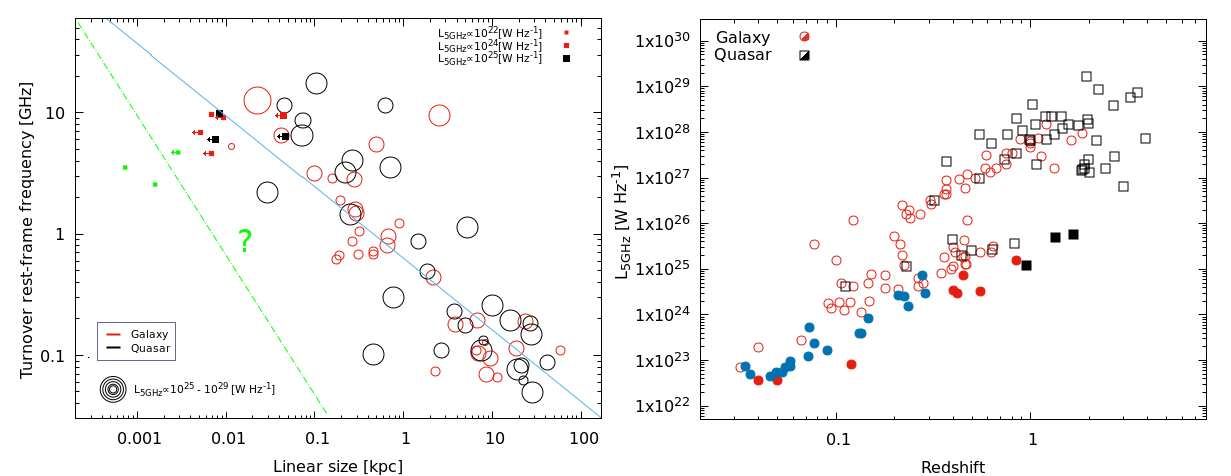}
    \caption{ {\bf Left}: the intrinsic turnover frequency vs. linear size relationship. The CSS/GPS comparison sample taken from the literature are indicated with 
    %\citet{Odea2}, \citet{Snellen}, \citet{devries}, \citet{Stranghellini}, and \citet{fanti}, 
circles and the transient GPS sources presented in this work are marked as squares (see \citet{me21} for more details). The sizes of the circles/squares correspond with the k-corrected radio luminosity at 5GHz, and arrows indicate maximum linear sizes for unresolved sources. The blue solid line indicates the linear relationship updated by \citet{Orienti}. The dashed green line indicates a separate development path for weak radio sources (green points) discussed in this article.
{\bf Right}: redshift vs. luminosity at 5GHz for the comparison sample combined with the low-luminosity CSS sources from \citet{MKB2010} (empty points), and transient objects presented in this work (filled points). The blue dots indicate the estimated position of the sources from the new VLASS sample. }
    \label{correlation}
\end{figure*}
\end{center}

All sources we discuss here are transient with respect to the FIRST survey (1995-2011) where they were undetected at the sensitivity level of $<$0.5 mJy at 1.4 GHz.
However, their detection in the CNSS survey (2012-2015) showed a significant increase in their brightness on a scale of several decades and therefore now they can be classified as radio-loud objects
($\rm log_{10}[L_{1.4GHz}/W~Hz^{-1}]>22.5$, Tab.\ref{basic}). 
The studied sample contains both quasars and galaxies in a wide range of redshift values ($\rm 0.04<z<1.7$) hence it is very diverse in terms of radio luminosity. However, their placement on a luminosity vs. redshift plane shows, that the group consist mainly of low-power sources with quasars being brighter than galaxies (Fig.\ref{correlation}, right).
%Their 3 GHz VLA flux density measurements with FIRST show that after the burst of radioactivity, the flux density stabilizes at an average level of a few to a dozen millijanskys. 
%Their redshift distribution is shown in Fig.\ref{correlation} with comparison to the sources from literature. Their placement on a luminosity vs. redshift plane shows, that the group consist mainly of low-power sources with quasars being brighter than galaxies. 

\begin{figure*}[t!]
    \centering
    \includegraphics[scale=1.29]{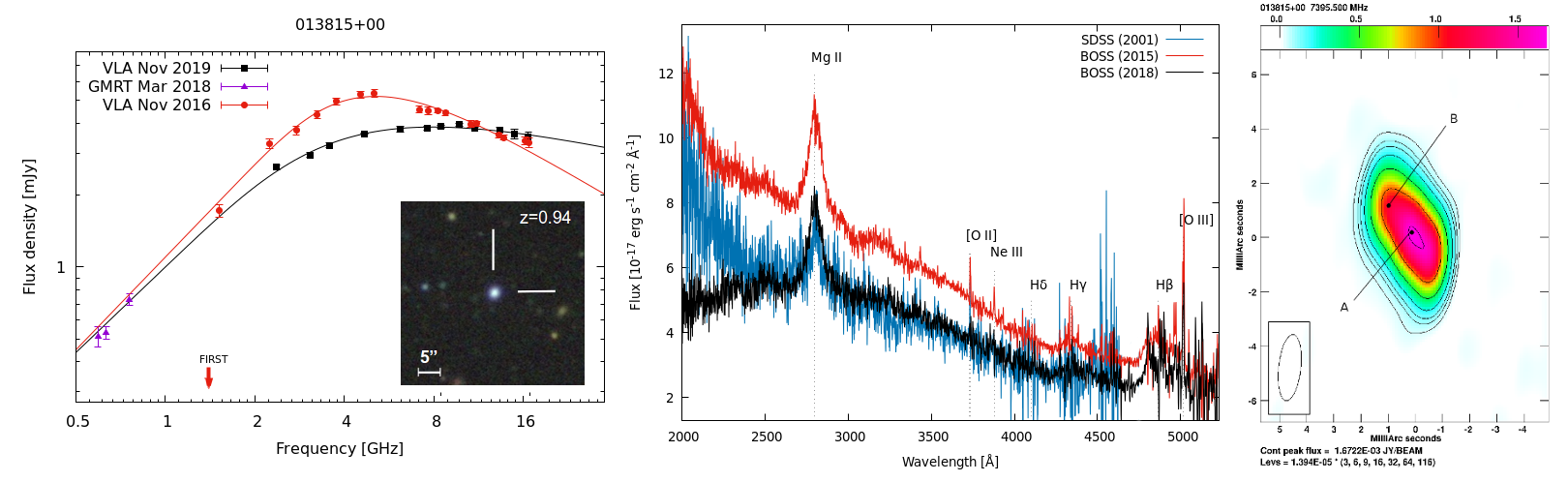}
    \caption{ Left: Radio spectrum of ``switched-on" radio AGN/quasar J013815+00 ($z=0.94$) from 0.5--18 GHz (red arrow is the upper limit from the FIRST survey from 1997). Inset shows the optical image cutout. Middle: The optical spectra across different epochs indicate accretion state changes over $\sim$years timescale. Right: The VLBA image reveals the young jet (core-jet morphology). Taken from Kunert-Bajraszewska et al.\ 2020.}
\label{013815}
\end{figure*}

The rich observation material (multi-frequency and mutli-epoch) that we managed to collect allowed for a detailed description of our sources and for tracing the changes that have occurred in them since the ignition of radio emission. 
And thus the VLA spectroscopic studies revealed that all the transient sources have convex spectra peaking in the range 
2$-$12 GHz at rest-frame. However, the radio spectra of quasars are changing rapidly with time. It is manifested by a flattening in the optically thin part of the spectrum. An example of such behavior are the changes seen in the spectrum of the quasar 013815$+$00 in Fig.\ref{013815}. As a result, after a few years of radio activity the quasars start to look like flat-spectrum objects, while galaxies keep their convex shape of the spectra.
%VLA spectroscopic studies also show that quasars are changing more rapidly with their spectra flattening quickly in the optically thin part, which may result in the fact that after a few years of radio activity they start to look like flat-spectrum objects and become hidden in this in this source population and therefore indistinguishable, while galaxies keep their convex shape of the spectra.
The phenomenon responsible for this characteristics is the birth of a new radio jet and the observed rapid changes are associated with its expansion and dissipation of energy. The differences in the behavior of the spectrum of galaxies and quasars over time are probably related to their different orientation in relation to the observer. And indeed the high-resolution 4.5 and 7.5 GHz VLBA images show the presence of small jets in case of some of our objects (see example in Fig.\ref{013815}). Nevertheless, these jets are not very prominent, and most of the objects remained unresolved in these observations. We argue that this may be due to the fact that the VLBA observations were carried out about 3 years after the ignition of radio activity in these sources, which in the case of weak jets may be enough time for the emission to fade out. 
%The convex shape, both in galaxies and the initial stages of quasar activity can be interpreted as a birth of new radio jet activity, while the rapid changes are associated with the evolution of the jet, which can indicate both its expansion or extinction.
%The phenomenon responsible for high variability, especially in the optically thin part of the spectrum, may also be the formation of new components in the source. Analysis of the optically thick part of the spectra shows much less significant variability, and the indices in this part of the spectra do not exceed the limit of 2.5 for a uniform source of synchrotron radiation, although several sources come very close to this limit, indicating a non homogeneous synchrotron component responsible for absorption in this part of the spectrum.\\  
%Maps obtained with VLBA show very compact, mostly pointilke morphology, slightly resolved in several cases, with a presence of only small jets or individual components, possibly due to a time discrepancy of about 3 years between the ignition of radio activity and the observations.\\

In order to test the hypothesis that our transient sources are the new born radio objects we placed them on the peak (turnover frequency) $\nu_p - l$ linear size plane (Fig.\ref{correlation}).
This relationship shows that the there is a continuous distribution of young AGNs with GPS sources evolving into CSS objects \citep{Odea97}.
%\citep{Odea,Orienti,Sotnikova}. This relationship suggests that the physical properties of the CSS and GPS sources are similar, and the only variable that depends directly on their size is their turnover frequency.
%Comparing the turnover frequency vs. linear size relation of our sources with the more powerful GPS/CSS sources from the literature shows, that in principle their 
We found that indeed the properties of our sources are similar to those of high-power GPS and CSS radio objects. They follow the established relationship quite accurately, with slight discrepancy visible in the case of the weakest sources. This may suggest a slower development of the source in terms of size and therefore the object will remain compact even for most of its life. We marked this as an alternative development path in Fig.\ref{correlation}(left). Nevertheless, this hypothesis requires further research on weaker radio objects in the early stages of evolution which is our research plan now (see the next section).\\  %indicating that insufficiently strong sources at some point of their development may ceise in their growth and remain compact for a longer time, or even permanently.\\
Attempts to detect our sample in the X-ray range resulted in three detections (two galaxies and one quasar) and the estimation of upper limits for six other sources. We then used the correlation between X-ray and radio luminosity to preliminarily classify them as low- and high-excitation radio galaxies. However, in general, the observational results of our sample in terms of X-ray show that the emission is weak, especially in the case of galaxies, and does not exceed the value of $\rm 10^{42}~erg~s^{-1}$. This, in turn, is consistent with the presence of a radiatively inefficient accretion onto a black hole with a mass within $10^6-10^8 M_{\odot}$. And indeed the black hole mass estimations we performed for our galaxies are in agreement with the suggested values \citep{me21}. In the case of quasars the black hole masses are higher, of the order of $10^9~M_{\odot}$. 

The optical spectroscopic data collected both from the SDSS archives and from our Southern African Large Telescope (SALT) project allowed for the estimation of a few more parameters of the discussed sources like bolometric luminosities, Eddington ratios and jet powers. Additionally, the detailed study of one of our quasars (Fig.\ref{013815}) led us to the conclusion that the ignition of radio activity
coincides with relatively small changes of bolometric
luminosity and hence Eddington ratio. We have observed that the burst of radio emission in 013815$+$00, recorded for the first time in December 2013, coincides with an increase in the brightness of the accretion disk (Fig.\ref{013815}, SDSS 2015). However, within the next two years, the disk brightness returns to its original state (SDSS 2018), which shows how fast changes occur in this new source. We estimated that the change in the bolometric luminosity of quasar 013815$+$00 was by a factor of just 1.4. Summarizing, a wide range of values of the physical parameters that we have obtained for our transient sources suggest that they belong to several different sub-classes of young radio objects.
%Since we have obtained a wide range of values for all of these parameters, we conclude that our sources belong to several different sub-classes of young radio objects.

%\begin{center}
%\end{center}

\section{New VLASS sample}
The objects described above constitute the first unbiased sample of 12 transient sources discovered in the CNSS survey \citep{MKB2020, me21}. However, the slow radio transient observations are continued thanks to the new ongoing Very Large Array (VLA) Sky Survey (VLASS; \citet{Lacy}). Very recently the detection of changing-state quasars has
been reported by \citet{Nyland} as a result of VLASS survey. %Their quasars show a large flux density increase in the S-band and convex radio spectra. 
However, 
%both research projects, this one of ours and this one of \citet{Nyland}, are only the beginning of the study of the slow radio variables and transient objects. 
a much larger population of such radio sources exists, and has been uncovered in the first epoch of VLASS survey, which we plan to explore further.

Our new sample consists of 24 newly-born candidates for short-lived radio AGN (Fig.\ref{correlation}, blue dots), undetected in the NVSS survey (1995), but discovered in VLASS to have brightened in the past $\sim$25 years. Selection criteria assumed point-like sources that were >8 mJy, coincident within 2 arcsec with the nuclei nearby galaxies having r<20 mag based on visual inspection of a Panoramic Survey Telescope and Rapid Response System (Pan-STARRS) optical images. 
The VLA archival sensitivity limits and new VLASS detections of these sources at 1.4 GHz implies spectral indices between 1.4 GHz and 3 GHz to be $\alpha>2$ ($S\propto\nu^{\alpha}$).
The sources are therefore either highly synchrotron self-absorbed GPS sources or are completely new bright sources in the radio sky. To better understand their nature, we have started a multi-frequency observation campaign for these objects with the use of radio and optical instruments.

%, the results of which will be published shortly.

%We have started a multi-frequency follow up campaign for these 24 transient galaxies.
%In either case, these sources give us an excellent opportunity to study a fairly uniform sample of young AGN in the "local" Universe.
%Therefore 

\section{Summary and future plans}\label{sec4}
We summarized here the first such extensive research of a sample of newly discovered radio sources. 
%Their radio emission has been 
They have been detected by CNSS survey between 2007 and 2013 on the level of a few mJy and higher. Based on their radio properties they can be classified as GPS objects, at least in the initial phase of radio activity.
They might have transitioned from radio-quiet to radio-loud
state either as a result of the increase in radio power
or its ignition.

However, thanks to the new and ongoing VLASS survey of the sky, new discoveries and studies of transient sources continues.
%The new project dedicated to tracking the transient phenomena, namely the VLASS survey, has revealed the existence of a large number of such events. 
%The first detection of transient quasars has been already reported \citep{Nyland}. 
%Our new sample consist of 
Our new research is now focused on 24 VLASS detections of candidates for young radio galaxies. These are nearby objects with a redshift $<$0.3.
Their radio luminosities at 3 GHz, $>10^{39}$ erg\ s$^{-1}$, suggest that they are most likely ``switched-on" AGN jets.
These sources give us an excellent opportunity to study a fairly uniform sample of young AGN in the "local" Universe  and will be published soon. 
%We have already started an extensive observation campaign of them with the use of radio and X-ray instruments.
%which means that they are in the radio-loud phase now.  
%presented the first original sample of 12 newly discovered sources with not previously detected radio activity, and its extensive research. 
%All sources are transient on a timescales of < 20 years and currently can be considered as radio-loud. They show compact morphology and convex spectra typical for GPS sources, with visible variability due to the evolution of their jets. Multi-frequency study of the sample shows a large diversity among the discussed young AGNs, including their radiative efficiency, accretion mode, physical parameters and the affiliation of their host galaxies to specific types of radio sources. The presented analysis is an important contribution to the study of the early stages of galaxy evolution and to revising past established models and their validity and universality. Further study assumes multi-frequency observations of a new, more extended sample of young AGNs and its analysis with improved methods and models that have been validated during the research presented here.\\
%More detailed study on the whole sample is presented in \citet{me21}, and the 013815$+$00 quasar has been discussed thoroughly by \citet{MKB2020}.

\section*{Acknowledgments}
%The National Radio Astronomy Observatory is a facility of the National Science Foundation operated under cooperative agreement by Associated Universities, Inc. We thank the staff of the VLA for carrying out these observations in their usual efficient manner.
MKB and AW acknowledge support from the  'National Science Centre, Poland' under grant no. 2017/26/E/ST9/00216.

%\subsection*{Author contributions}
%This is an author contribution text. This is an author contribution text. This is an author contribution text.  
%\subsection*{Financial disclosure}
%???
%\subsection*{Conflict of interest}
%The authors declare no potential conflict of interests.

%\nocite{*}% Show all bib entries - both cited and uncited; comment this line to view only cited bib entries;
\bibliography{main}%

%\section*{Author Biography}

%\begin{biography}{\includegraphics[width=60pt,height=70pt,draft]{empty}}{\textbf{Aleksandra Wo\lowska} is completing her doctoral dissertation at the Faculty of Physics, Astronomy and Informatics at the Nicolaus Copernicus University in Toruń, Poland. Her main topic of research is extragalactic sources, in particular the characteristics and evolution of the youngest objects.}
%\end{biography}

\end{document}